\documentstyle[12pt,graphicx,amssymb]{article}

\begin{document}

\title{Low-density series expansions for directed percolation I:\\
A new efficient algorithm with applications to the square lattice}

\author{Iwan Jensen \thanks{e-mail: I.Jensen@ms.unimelb.edu.au} \\
Department of Mathematics and Statistics, \\
University of Melbourne, Parkville, Victoria 3052, Australia}

\maketitle

\begin{abstract}
A new algorithm for the derivation of low-density series for percolation 
on directed lattices is introduced and applied to the square lattice
bond and site problems. Numerical evidence shows that the 
computational complexity grows exponentially, but with a growth factor 
$\lambda < \protect{\sqrt[8]{2}}$, which is much smaller than the growth 
factor $\lambda = \protect{\sqrt[4]{2}}$ of the previous best algorithm. 
For bond (site) percolation on the directed square lattice the series has 
been extended to order 171 (158). Analysis of the series yields
sharper estimates of the critical points and exponents.
\end{abstract}


\section{Introduction \label{sec:intro}}

Directed percolation (DP) \cite{DPreview,PercTh} can be thought of as a 
purely geometric model in which the bonds or sites of a hyper-cubic 
lattice ${\Bbb Z}^d$ are either present with probability $p$ or 
absent (with probability $q=1-p$). Unless otherwise specified I 
shall be looking at bond percolation on the directed square 
lattice. As in ordinary percolation one might be interested in 
the average cluster-size $S(p)$. The difference is that in 
DP connections are only allowed along a preferred direction given 
by an orientation of the edges. Fig.~\ref{fig:oriented} shows 
the part of the square lattice which can be reached from the 
origin using no more than five steps. Apart from the inherent
theoretical interest DP has been associated with a wide 
variety of physical phenomena. In {\em static} interpretations,
the preferred direction is spatial, and DP could model gravity
driven percolation of fluid through porous rock with
a certain fraction of the channels blocked \cite{DE},
crack propagation \cite{Kertesz} or electric current
in a diluted diode network \cite{Redner}. In {\em dynamical}
interpretations, the preferred direction is time and one 
realization is as a model of an epidemic without immunization. DP type 
transitions are also encountered in many other situations including
Reggeon field theory \cite{RFT}, 
chemical reactions \cite{chem}, heterogeneous catalysis 
and other surface reactions \cite{catalysis}, 
self-organized criticality \cite{SOC}, and
even galactic evolution \cite{galactic}. 

Domany and Kinzel \cite{DKmodel} demonstrated that bond percolation 
on the directed square lattice is a special case of a one-dimensional 
stochastic cellular automaton. The evolution of the model is governed 
by the transition probabilities $W(\sigma_x | \sigma_l,\sigma_r)$,
with $\sigma_i = 1$ if site $i$ is occupied and 0 otherwise. It is the
probability of the site $x$ being in state $\sigma_x$ at time $t$ given 
that the sites $x-1$ and $x+1$ at time $t-1$ were in states $\sigma_l$ 
and $\sigma_r$, respectively. One has a free hand in choosing the 
transition probabilities as long as one respects conservation of 
probability, $P(1|\sigma_l,\sigma_r)=1-P(0|\sigma_l,\sigma_r)$.
Bond percolation corresponds to the choice:

\begin{equation}
W(0| \sigma_l,\sigma_r) = (1-p)^{\sigma_l+\sigma_r}  
\label{eq:sqbrancprob}
\end{equation}

\noindent
Note that while one talks of occupied {\em sites} in the cellular
automata language, the choice of $W$ above does indeed correspond
to {\em bond} percolation in the static version, as can easily be
verified by explicitly looking at all possible configurations and
their associated probabilities.

The behavior of the model is controlled by the branching
probability (or density of bonds) $p$. When $p$ is smaller 
than a critical value $p_c$ all clusters remain finite. Above $p_c$
there is a non-zero probability of finding an infinite cluster and
$S(p)$ diverges as $p \rightarrow p_c^-$.
In the low-density phase ($p<p_c$) many quantities of interest 
can be derived from the pair-connectedness $C_{x,t}(p)$, which
is the probability that the site $x$ is occupied at time
$t$ given that the origin was occupied at $t=0$. The moments
of the pair-connectedness may be written as

\begin{equation}
\mu_{n,m}(p) = \sum_{t=0}^{\infty}\sum_{x} x^nt^m C_{x,t}(p).
\label{eq:moments}
\end{equation}

\noindent
In particular the average cluster size $S(p)=\mu_{0,0}(p)$.
Due to symmetry, moments involving odd powers of $x$ vanish.
The remaining moments diverge as $p$ approaches the critical point
from below:

\begin{equation}
\mu_{n,m}(p) \propto (p_c-p)^{-(\gamma+n\nu_{\perp}+m\nu_{\parallel})},
 \;\;\;\; p \rightarrow p_c^-.
\end{equation}

For isotropic percolation many exact results are known for
two-dimensional systems, e.g., the critical exponents are known
exactly from arguments based on conformal field theory \cite{Cardy87} 
and for some lattices the critical point is 
known exactly \cite{SE}. Isotropic bond percolation is also related 
to the $q \rightarrow 1$ limit of the Potts model \cite{WUrev,RCM}. 
For directed percolation no such exact results are known, except for
special limited versions such as compact directed percolation 
\cite{EssamCDP}. Though it is worth mentioning that directed bond
percolation is related to the $q \rightarrow 1$ limit of the 
{\em chiral} Potts model \cite{AEchiral} and to the $m \rightarrow 0$ 
limit of the $m$-friendly walker problem \cite{Tsuchiya,Cardy99}.
So in order to study directed percolation one has to 
resort to numerical methods. For many problems
the method of series expansions is by far the most powerful method
of approximation. For other problems Monte Carlo methods are superior.
For the study of DP on two-dimensional lattices, series analysis is 
undoubtedly the most appropriate choice. This method consists of 
calculating the first coefficients in the expansion of $\mu_{n,m}(p)$. 
Given such a series, using the numerical technique known as 
differential approximants \cite{Guttmann89}, 
highly accurate estimates can frequently be obtained for  the 
critical point and exponents, as well as the location and critical 
exponents of possible non-physical singularities.

The difficulty in the enumeration of most interesting lattice problems
is that, computationally, they are of exponential complexity.  Initial 
efforts at computer enumeration of square lattice directed percolation 
were based on direct counting. The computational complexity is proportional 
to $\lambda_1^n,$ where $n$ is the number of terms in the series, and 
$\lambda_1 =3$, is the connective constant for directed 
lattice animals on the square lattice \cite{DPB}.  A dramatic improvement 
was achieved by the transfer matrix technique of Blease who devised an 
algorithm with a complexity, which is proportional to $\lambda_2^n,$ where
$\lambda_2 =  \sqrt{2} \approx 1.414.$ This work resulted in a derivation
of the first 21 terms in the low-density series on the square lattice. 
Over the years further significant progress was obtained by Essam and 
co-workers who supplemented the transfer matrix technique by
a weak subgraph expansion and obtained first 25 terms \cite{DBE}  and
some years later 35 terms \cite{EDBAB}. Note that while the  weak subgraph 
expansion enables one to derive more terms it does not improve the
computational complexity.
Eventually they devised a resummation technique \cite{EGDB} which was used
to derive the so-called non-nodal graph expansion \cite{BE}. This in turn 
enabled them to obtain twice as many terms \cite{EGDB} as could be obtained 
by the bare algorithm of Blease, resulting in a series of 49 terms. 
So this approach has a complexity, which is proportional 
to $\lambda_3^n,$ where $\lambda_3 =  \sqrt[4]{2} \approx 1.189.$ 
This method was later supplemented by an extension method 
\cite{Jensen96a}, originally suggested by work of Baxter and Guttmann 
\cite{BG}, based on predicting correction terms from successive  
calculations on finite lattices of increasing size. Again this extension 
method does not improve the computational complexity but it did, in
conjunction with several years of improvements to computer technology, 
result in an extension of the series to 112 terms.  
In this paper I shall describe a new algorithm based on a direct calculation 
of the non-nodal graph expansion counting only graphs with non-zero 
weights. This has reduced both time and storage  requirements by virtue 
of a complexity proportional to $\lambda_4^n,$ where the numerical evidence 
suggest that $\lambda_4 \leq \sqrt[8]{2} \approx 1.091$. The series
has now been calculated to 171 terms using essentially the same
computational resources used earlier to derive 112 terms, but without 
the need for a cumbersome and complicated extension procedure. 
The series for the directed square lattice site problem has been 
extended to 158 terms from the previous best of 106 terms.

In the next section I will briefly review the previous best method for
deriving low-density series expansion for square lattice directed 
percolation. In section~\ref{sec:algo} I give a detailed description and 
empirical analysis of the computational complexity of the improved 
algorithm. The results of the analysis of the series are presented 
in Section~\ref{sec:analysis}.

\section{Series expansions \label{sec:serexp}}

From Eq.~(\ref{eq:moments}) it follows that the first and 
second moments can be derived from the quantities

\begin{equation} \label{eq:SXdef}
S(t) = \sum_{x} C_{x,t}(p)\;\;\; \mbox{ and }\;\;\; 
X(t) = \sum_{x} x^2 C_{x,t}(p).
\end{equation}

\noindent
$S(t)$ and $X(t)$ are polynomials in $p$ obtained by summing the
pair-connectedness over all lattice sites whose parallel
distance from the origin is $t$. It has been shown \cite{AE} that 
the pair-connectedness can be expressed as a sum over all graphs 
(or finite clusters) formed by taking unions
of directed paths connecting the origin to the site $(x,t)$,
\begin{equation}\label{eq:pcg}
C_{x,t}(p)= \sum_{g} d(g)p^{|g|},
\end{equation}
where $|g|$ is the number of bonds in
the graph $g$. Any directed path to a site whose parallel distance 
from the origin is $t$ contains at least $t$ bonds. 
From this it follows
that if $S(t)$ and $X(t)$ have been calculated for $t \leq t_{\rm max}$
then one can determine the moments to order $t_{\rm max}$. One can
however do much better as demonstrated by Essam et al.
\cite{EGDB}. They used a non-nodal graph expansion, based on
work by Bhatti and Essam \cite{BE}, to extend the series
to order $2t_{\rm max}+1$.
The non-nodal graph expansion has been described in detail
in \cite{EGDB} and here I will only summarize the main points
and introduce some notation. A graph $g$ is nodal if there is
a point (other than the terminal point) through which all paths
pass. It is clear that each such nodal point effectively works as
a new origin for the cluster growth. This is the essential idea
behind the non-nodal graph expansion. Let $S^N(t)$ and $X^N(t)$
be the contribution from non-nodal graphs to $S(t)$ and $X(t)$, 
respectively. The non-nodal expansions are obtained recursively from 
the polynomials $S(t)$ and $X(t)$ \cite{EGDB}. Next one forms the 
moments $S^N$, $\mu_{0,1}^N$, $\mu_{0,2}^N$, and $\mu_{2,0}^N$  
of non-nodal contributions equivalent to  Eq.~(\ref{eq:moments}). 
The final series are obtained from the formulas
\begin{eqnarray}
  S & = & 1/(1-S^N)  \\
  \mu_{0,1} & = & \mu_{0,1}^NS^2 \\
  \mu_{0,2} & = & [\mu_{0,2}^N+2(\mu_{0,1}^N)^2S]S^2 \\
  \mu_{2,0} & = & \mu_{2,0}^NS^2.
\end{eqnarray}

Further extensions of the series can 
be obtained by using a procedure similar to that  
of Baxter and Guttmann \cite{BG}. One looks at
correction terms to the series and tries to identify extrapolation
formulas for the first $n_r$ correction terms allowing one to derive
a further $n_r$ series terms correctly. Details of this procedure
can be found in \cite{Jensen96a}.

\begin{figure}
\begin{center}
\includegraphics{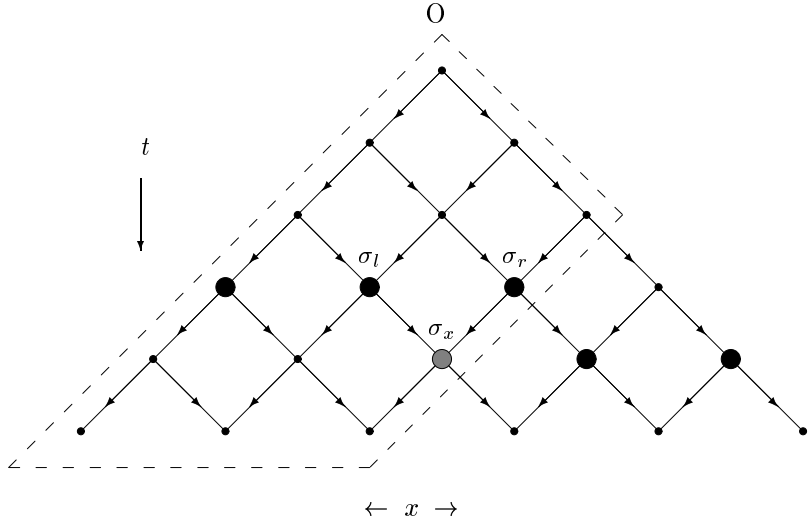}
\end{center}

\caption{\label{fig:oriented} The directed square 
lattice with orientation given by the arrows.}

\end{figure}

The pair-connectedness can be calculated \cite{Jensen96a} using a 
transfer-matrix technique based on the Domany-Kinzel model. The 
probability of finding a given configuration of sites at time $t$ was 
calculated by moving a boundary through the lattice one site at a time.
Fig.~\ref{fig:oriented} shows how the boundary (marked by large filled 
circles) is moved in order to pick up the weight associated with a given 
`face' of the lattice at a position $x$ along the boundary line.
At any given stage this line cuts through a number of, say $k$,
lattice sites thus leading to a total of $2^k$ possible configurations
along this line. Configurations along the boundary line are trivially
represented as binary numbers, and the probability of each 
configuration is given by a truncated polynomial in $p$. Let
$S0=(\sigma_1,....,\sigma_{x-1},0,\sigma_{x+1},...,\sigma_k)$ be the
configuration of sites along the boundary with 0 at position $x$ and
similarly $S1=(\sigma_1,....,\sigma_{x-1},1,\sigma_{x+1},...,\sigma_k)$
the configuration with 1 at position $x$. Then in moving the boundary at
the  $x'$th site the polynomials are updated as follows 
\begin{eqnarray*}
P(S0) & = & W(0|0,\sigma_l)P(S0)+W(0|1,\sigma_l)P(S1),  \\
P(S1) & = & W(1|0,\sigma_l)P(S0)+W(1|1,\sigma_l)P(S1).
\end{eqnarray*}

The pair-connectedness is obviously symmetrical in $x$,
$C_{x,t}(p) = C_{-x,t}(p)$, so it suffices to calculate the
pair-connectedness for $x \geq 0$. More importantly, due to the
directedness of the lattices, if one looks at sites $(x,t)$ with
$x \geq 0$ they can never be reached from points $(x',t')$ in the 
part of the lattice for which 
$t'> \lfloor t/2\rfloor, \; x' < -\lfloor t/2\rfloor$. 
In Fig.~\ref{fig:oriented} the part of the lattice needed in order
to calculate $C(x,t)$ up to $t=5$ is enclosed by the dashed lines and
the two sites on the boundary-line to the right of the dashed line
can be disregarded. In more general terms this means that the 
pair-connectedness at points a distance $t$ from the origin can 
be calculated using a boundary which cuts through at most 
$\lfloor t/2\rfloor+1$ sites. Thus the memory (and time) required
to calculate $S(t)$ and $X(t)$ grows like $2^{\lfloor t/2\rfloor+1}$.
Using the non-nodal expansion it follows that the time and memory 
required to calculate the series to a given order $n$ grow like
$2^{n/4}$, and thus that the computational complexity of this
algorithm is exponential with growth-factor $\lambda = \sqrt[4]{2}$.
In \cite{Jensen96a} this calculation was performed up to 
$t_{\rm max} =47$. By looking at correction terms the series for $S(p)$, 
$\mu_{0,1}(p)$ and $\mu_{0,2}(p)$ were extended to order 112 and 
the series for $\mu_{2,0}(p)$ to order 111.

\section{The new algorithm \label{sec:algo}}

\begin{figure}
\begin{center}
\includegraphics{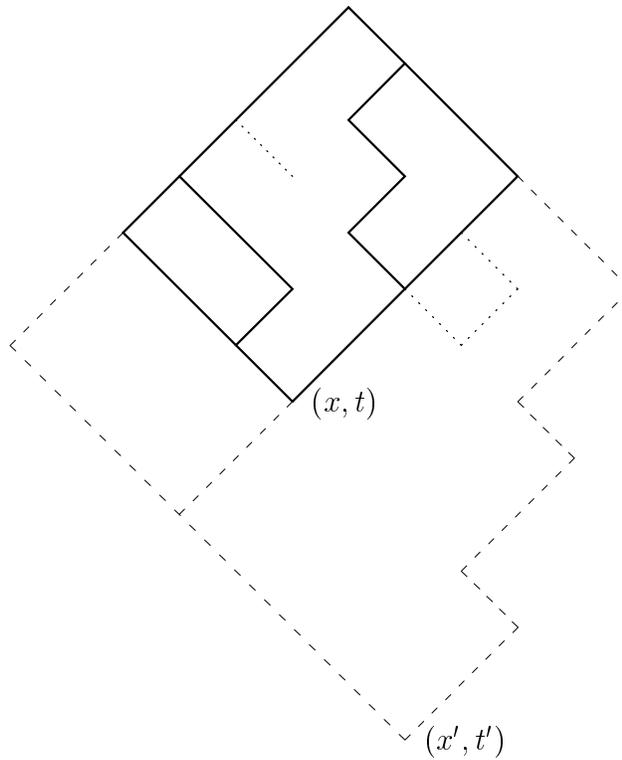}
\end{center}

\caption{\label{fig:contribute} Illustration of graphs which
contribute to $C(x,t)$ (solid lines), $C(x',t')$ (solid+dashed lines) 
and paths which are {\em not} allowed (dotted lines).}

\end{figure}

In the following I shall describe a new algorithm based on calculating
the pair-connectedness using the expression in terms of contributing 
graphs given in Eq.~(\ref{eq:pcg}).  The algorithm directly calculates 
the non-nodal contributions and the numerical evidence shows that 
asymptotically the computational complexity has a
growth-factor $\lambda \leq \sqrt[8]{2}$. As already noted, the directed 
lattice weights $d(g)$ are non-zero if and only if the graph $g$
is a union of paths from the origin to $(x,t)$. The graphs with
non-zero weights form a very limited set of all possible graphs
passing through $(x,t)$. The restriction to unions of paths is 
very strong and one immediate consequence is that graphs with
dangling/dead-end parts make no contribution. Likewise a graph
without dangling parts makes a contribution to $C(x,t)$ only 
if it terminates exactly at $(x,t)$. It could of course contribute
at a later stage when the various branches join. 
Fig.~\ref{fig:contribute} illustrates these points. The thick solid
lines shows an example of a graph contributing to $C(x,t)$. The 
dotted lines illustrate dangling ends which are {\em not} allowed,
and the thin dashed lines show branches which would have to
be joined at a later stage in order for this extended graph to 
make a contribution to $C(x',t')$.

The directed weights have the further properties \cite{AE}
\begin{equation}\label{eq:dwprop}
d(g) = d(g')d(g''), \;\; d(g) = 0,\; \pm 1.
\end{equation}

\noindent
where $g'$ and $g''$ are subgraphs of $g$. So the $d$-weights factorize
and for contributing graphs are either 1 or $-1$. Furthermore, for
contributing graphs $d(g)=(-1)^{t(g)+1}$, where $t(g)$ is the
number of independent paths from the origin to the end point of the
graph $g$. The calculation of $d(g)$ is in principle quite complicated 
were it not for the factorization property and the fact that factors 
of $-1$ are picked up each time two paths join (or alternatively
when one path splits into two paths). Indeed one sees that $t(g)$
is simply the number of times a path splits into two, which of course
equals the number of times two paths join to form a single path. This 
means that once again the pair-connectedness can be calculated via a 
transfer-matrix type algorithm by moving a boundary line through 
the lattice one row at a time with each row built up one site at a time. 
The sum over all contributing graphs is calculated as one goes along. 

\begin{figure}
\begin{center}
\includegraphics{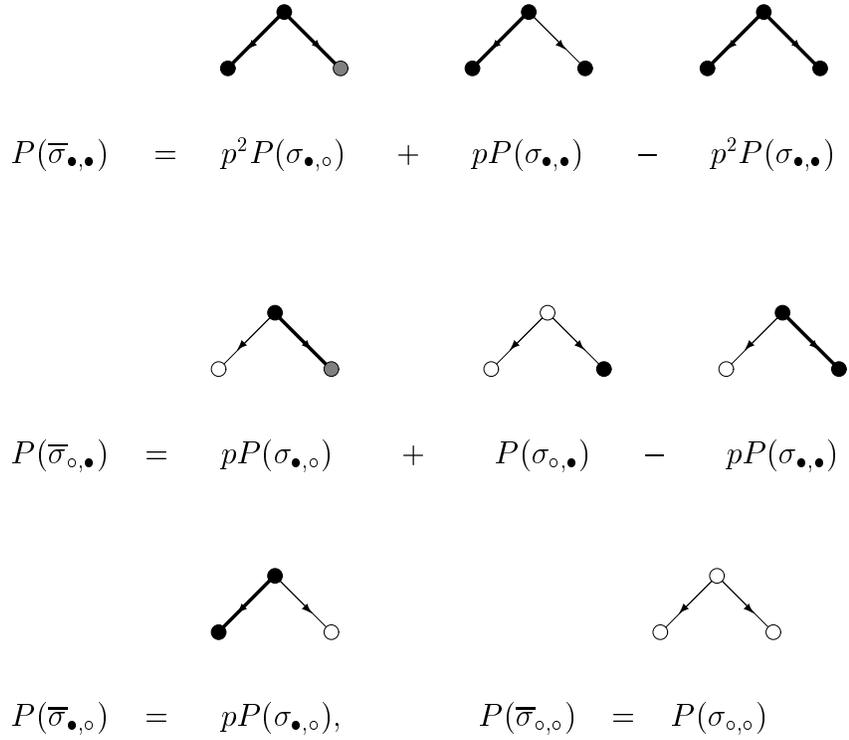}
\end{center}

\caption{\label{fig:newupdate} Pictorial representation of the
rules for updating boundary polynomials in the new algorithm.}

\end{figure}

In Fig.~\ref{fig:newupdate} I have given a pictorial representation
of how the boundary line polynomials are updated by the new algorithm.
First it should be noted that there are two states per site
with the following prescription: $\sigma_{i}= 1$ if a bond 
has been inserted along an edge from the row above, and $\sigma_{i}= 0$ 
otherwise. We are looking at a move similar to that in 
Fig.~\ref{fig:oriented}, though in this case
the sites involved in the move are the one on the top and its
neighbours in the row below.  I shall refer to the configuration
after the move as the `target' configuration  and it accounts for the 
state of the sites on the bottom row. The configurations prior to the
move, given by the state of the top and bottom right sites, shall be
called `source' states. The state of sites away from the kink in 
the boundary does not influence the updating.  In 
Fig.~\ref{fig:newupdate} sites with incoming bonds are marked with 
a filled circle, those with open circles have no incoming bonds, 
sites marked by a shaded circle have incoming bonds in the 
target state but not in the source state, and bonds are marked
by thick lines. Note that the avoidance of
dangling ends is easily achieved by ensuring that sites with incoming
bonds have at least one outgoing bond. In the first equation the 
target configuration, $\overline{\sigma}_{\bullet,\bullet}$, 
has bonds entering both sites and is fed by the source configurations
$\sigma_{\bullet,\circ}$ and $\sigma_{\bullet,\bullet}$. In moving the 
boundary from the top site, bonds can be inserted 
to the left and/or right. The source state $\sigma_{\bullet,\circ}$ 
has no bonds entering the left site so both bonds have to be added
with an associated weight $p^2$. From the source state 
$\sigma_{\bullet,\bullet}$ a bond has to be inserted on the left edge, 
but on the right edge a bond can be either absent, in which case an
associated weight $p$ is required, or a bond can be added giving a 
weight $-p^2$ because two bonds were inserted and two paths join on 
the site to the right. Similar considerations lead to the other equations.

Limiting the calculation to non-nodal contributions is very simple; 
whenever the boundary line reaches the horizontal position all one need do 
is set to zero the polynomials of states with a single incoming bond, 
i.e., states represented by integers of the form $2^k$.
This obviously ensures that configurations  with a nodal point
are deleted from the calculation. The pair-connectedness at the
following time can be calculated from the states with incoming
bonds at nearest neighbour sites and no incoming bonds on any other sites. 

In a calculation to a given order $n$ we need to calculate the
non-nodal contributions for all $t \leq t_{\rm max}=n/2$. For a
given $t' < t$ the possible configurations along the boundary line is
limited by constraints arising from the facts that graphs have
to terminate at level $t$ and have no dangling parts. As mentioned,
the ``no dangling parts'' restraint is equivalent to demanding that
sites with incoming bonds also have outgoing bonds. Therefore
a configuration for which the maximal separation between sites
with incoming bonds is $r$ will take at least another $r-1$ steps
before collapsing to a configuration with a pair of nearest neighbour
sites with incoming bonds. Consequently if $t'+r >  t$ that configuration
makes no contribution to $C(x,t)$ for any $x$ and can be discarded. 
Furthermore the minimum number of bonds needed to build up the internal 
structure of the configuration (that is the occupied sites except for the 
left- and right-most sites) will have to be inserted during the collapse 
of the configuration.  Since we are interested only in non-nodal 
graphs we further know that we have to insert at least $2t$ bonds to 
get from the origin to a point $(x,t)$. It is easy to calculate the 
minimum order, $N_{\rm min}$, of the boundary polynomial as the
configuration is built up, but in so doing we have also counted $2t'$ 
bonds coming from the paths leading to the left- and right-most sites,
respectively. So the minimal order to which a configuration contributes, 
$N_{\rm cont}$, from row $t'$ is

\begin{equation}\label{eq:sqbcont}
N_{\rm cont}=2N_{\rm min}+2t-4t'.
\end{equation}

\noindent
If $N_{\rm cont} > n$ the configuration can 
be discarded since it will only contribute at an order exceeding that to 
which we want to carry out our calculation. Further memory savings are 
obtained by observing that in calculating  $C(x,t)$ we know that the 
non-nodal graphs have at least $2t$ bonds, so we need only store 
$n-2t$ coefficients, and when the boundary is moved from one row to 
the next we discard the two lowest order terms in the boundary 
polynomials since they are all zero.

The algorithm for calculating the series for directed site percolation
is very similar, but of course we have different rules for updating
the boundary polynomials. Though it should be noted that the meaning
of `0's and `1's in the encoding of configurations is changed slightly
to signify unoccupied and occupied sites respectively.
The updating rules are easy to derive from those
for the bond problem. In fact all we need to do is take the rules
depicted in Fig.~\ref{fig:newupdate} and wherever we have inserted a
bond on the right to a site already occupied the weight associated
with that possibility is simply divided by $p$ because in {\em site} 
percolation we cannot reoccupy a site already occupied. This leads
to the following update rules for the site problem:

\begin{eqnarray}
P({\overline{\sigma}}_{\bullet,\bullet})& = &
p^2 P(\sigma_{\bullet,\circ}), \nonumber \\
P({\overline{\sigma}}_{\circ,\bullet})& = & p P(\sigma_{\bullet,\circ}) + 
P(\sigma_{\circ,\bullet}) - P(\sigma_{\bullet,\bullet}),\nonumber  \\
P({\overline{\sigma}}_{\bullet,\circ}) & = &
p P(\sigma_{\bullet,\circ}), \\
P({\overline{\sigma}}_{\circ,\circ})& = & P(\sigma_{\circ,\circ}). \nonumber 
\end{eqnarray}

\noindent
The only other change is that the formula for the minimal order
of a contributing configuration is changed a little, becoming

\begin{equation}
N_{\rm cont}=2N_{\rm min}+2t-N_{\rm occ}-4t'
\end{equation}

\noindent
where we have subtracted the number of occupied sites, $N_{\rm occ}$, in 
the configuration (otherwise they would have been counted twice from the
first factor).

\begin{figure}
\begin{center}
\includegraphics{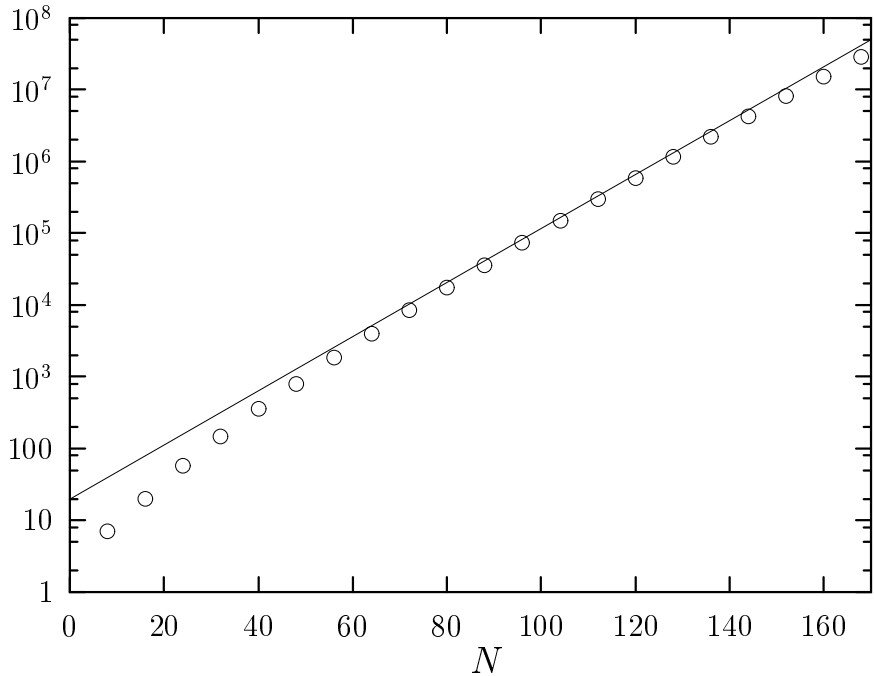}
\end{center}

\caption{\label{fig:memuse} The maximal number of contributing
configurations in a calculation to order $N$.
The straight line is a pure exponential $\propto \lambda^N$ with growth 
factor $\lambda = \protect{\sqrt[8]{2}}$.}

\end{figure}

In Fig.~\ref{fig:memuse} I have plotted the number of distinct configurations
which make a contribution in a calculation of the square lattice bond series 
to order $n$. As can be seen the numerical evidence clearly shows that the 
asymptotic growth in the number of configurations  is exponential with a 
growth factor $\lambda \leq \sqrt[8]{2}$. As already noted, this is a very
substantial improvement on the previous best algorithm which had a
growth factor $\lambda = \sqrt[4]{2}$. In plain words this means that
whenever computer memory is doubled one can derive an additional 8 terms
with the new algorithm as compared to an additional 4 terms with the previous
best algorithm. Using all the memory minimization tricks mentioned above it 
is possible to derive the series for moments of the pair-connectedness
to order 171 with a maximum of approximately 5.5Gb of memory. The old 
algorithm would have required more than 250~000 times as much memory and 
even if the extension procedure was used to derive say the last 20 terms 
correctly more than 10~000 times as much memory would have been required.
For directed site percolation the series has been derived to
order 158 using similar computational resources.

Finally a few remarks of a more technical nature. 
The number of contributing configurations becomes very sparse
in the total set of possible states along the boundary line and
as is standard in such cases one uses a hash-addressing scheme
\cite{Mehlhorn}. It is probably also worth mentioning that since
the update involves only two nearest neighbour sites and does not 
depend on the state of other sites along the boundary, this algorithm 
lends itself very naturally to parallel computing. 
Since the integer coefficients occurring in the series expansion 
become very large, the calculation was performed using modular 
arithmetic \cite{Knuth}. This involves performing the calculation modulo 
various prime numbers $p_i$ and then reconstructing the full integer
coefficients at the end. In order to save memory I used primes of the form 
$p_i=2^{15}-r_i$ so that the residues of the coefficients in the polynomials 
can be stored using 16-bit integers. The Chinese remainder theorem
ensures that any integer has a unique representation in
terms of residues. If the largest absolute values occurring
in the final expansion is $m$, then we have to use a number of
primes $k$ such that $p_1p_2\cdots p_k/2 > m$. Note that
it is not necessary to be able to uniquely reproduce the
intermediate values, which in some cases can be much larger than 
the final ones. Up to 12 primes were needed to represent the 
coefficients correctly.

The updating rules given above are quite general and can be used for 
calculating the pair-connectedness of percolation problems on the 
directed square lattice in many other cases. One example is that of a 
lattice with an impenetrable wall \cite{EGJT}, as shall be demonstrated 
in a forthcoming article, though in this case the 
calculation is not confined to non-nodal graphs. Other applications 
include directed percolation with temporal or spatial disorder in which
the branching probability $p$ depends on $t$ or $x$ \cite{Jensen96b}. The 
basis for the algorithm used in this paper is Eq.~(\ref{eq:pcg}). It is a 
very general expression for the pair connectedness which in fact should 
hold on any directed lattice and also for more complicated processes. Thus 
the work started in this paper can be generalized to other planar 
lattices or to higher dimensional lattices, though naturally the 
updating rules and other details of the actual implementation will vary
from application to application. Another very interesting 
possibility is the application to unidirectionally coupled directed 
percolation \cite{THH98,GHHT99,Janssen99}. In this case one studies a
hierarchy of identical directed percolation processes. The 0'th level 
process is just an ordinary directed percolation process.  The $k$'th 
level process is a directed percolation process evolving according to 
the usual rules. But in addition sites $(x,t)$ may become occupied at some
given rate if the corresponding site was occupied at level $(k-1)$. 
Studies of this system showed that the exponents depend on $k$, in
essence showing that as $k$ is increased percolation becomes easier.
By truncating the hierarchy at some low value of $k$ it should be
possible to generalize the algorithm to  study this new and very
interesting problem. 

\section{Analysis of series \label{sec:analysis}}

The series for moments of the pair-connectedness were analyzed 
using differential approximants. A comprehensive review of
these and other techniques for series analysis may be found in
\cite{Guttmann89}.  This allows one to locate the critical point 
and estimate the associated critical exponents fairly accurately, 
even in cases  where there are additional non-physical singularities.  
Here it suffices to say that a $K$th-order
differential approximant to a function $f$, for which one has derived
a series expansion, is formed by matching the coefficients in the
polynomials $Q_i$ and $P$ of order $N_i$ and $L$, respective,
so that the solution to the inhomogeneous differential equation
\begin{equation}\label{eq:diffapp}
\sum_{i=0}^K Q_{i}(x)(x\frac{\mbox{d}}{\mbox{d}x})^i \tilde{f}(x) = P(x)
\end{equation}
agrees with the first series coefficients of $f$. The equations are
readily solved as long as the total number of unknown coefficients in
the polynomials is smaller than the order of the series $n$.
The possible singularities of the series appear as
the zeros $x_i$ of the polynomial $Q_K$ and the associated critical
exponent $\lambda_i$ is estimated from the indicial equation

$$
\lambda_i=K-1-\frac{Q_{K-1}(x_i)}{x_iQ_K ' (x_i)}.
$$
The physical critical point is generally the first singularity on
the positive real axis.

\subsection{The bond series}

In this section I will give a detailed account of the results of the
analysis of the square bond series. The analysis of the square site is 
described in the following section. In addition to the
moment series I have also analyzed the series
$\mu_{0,2}(p)/\mu_{0,1}(p) \sim (p_c-p)^{-\nu_{\parallel}}$ and  
$\mu_{2,0}(p)\mu_{0,2}(p)/(\mu_{0,1}(p))^2\sim (p_c-p)^{-2\nu_{\perp}}$.

As one increases the order $N_i$ and $L$ of the polynomials in 
Eq.~\ref{eq:diffapp} and thus the number of terms used to form the 
differential approximants, one would generally expect to obtain
more accurate estimates for the critical parameters. Likewise
one often finds that the estimates show some dependence on the
order $K$ of the differential approximant and/or the order $L$
of the inhomogeneous polynomial. In previous work \cite{Jensen96a}
it was observed that for some series the estimates from first-order
differential approximants showed a marked change with increasing
$L$. Analysis of the longer series of course confirm this and
also that some series (in particular $\mu_{0,1}(p)$ and $\mu_{0,2}(p)$)
show a certain drift as $K$ is increased. But over all the estimates
from the various series are exceptionally well converged and
for $K\geq 2$ show little if any change as $L$ is changed or $K$
increased. 

\begin{figure}
\begin{center}
\includegraphics{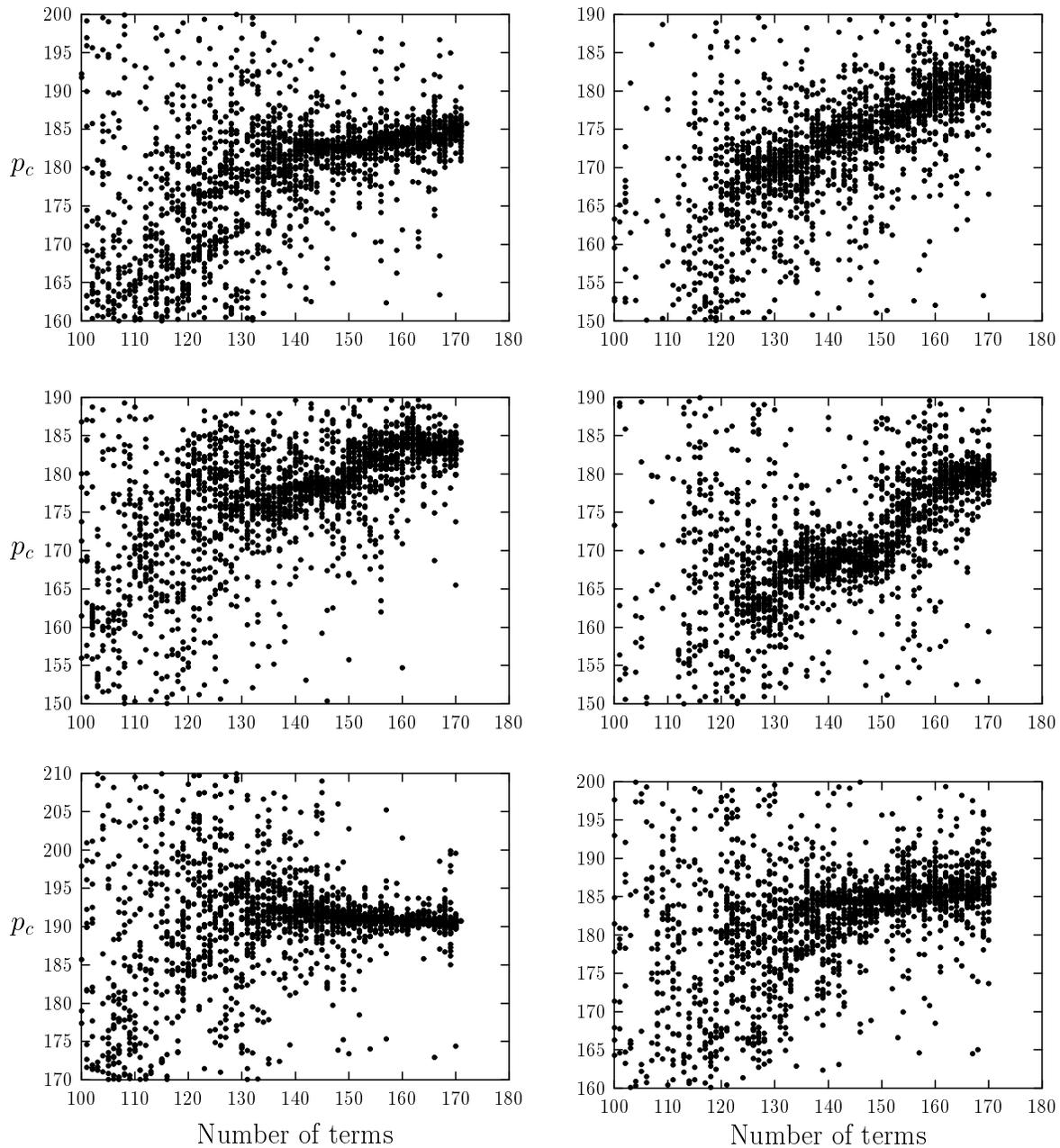}
\end{center}

\caption{\label{fig:ntcrp} Estimates of the critical point $p_c$
obtained from third-order differential approximants vs 
the number of terms used by the approximant. 
Numbers along the $y$-axis are all preceded by 0.644700.
Shown are (from left to right and top to bottom) estimates from
the series $S(p)$, $\mu_{0,1}(p)$, $\mu_{2,0}(p)$, $\mu_{0,2}(p)$,
$\mu_{2,0}(p)\mu_{0,2}(p)/(\mu_{0,1}(p))^2$,
and  $\mu_{0,2}(p)/\mu_{0,1}(p)$.}

\end{figure}

\begin{figure}
\begin{center}
\includegraphics{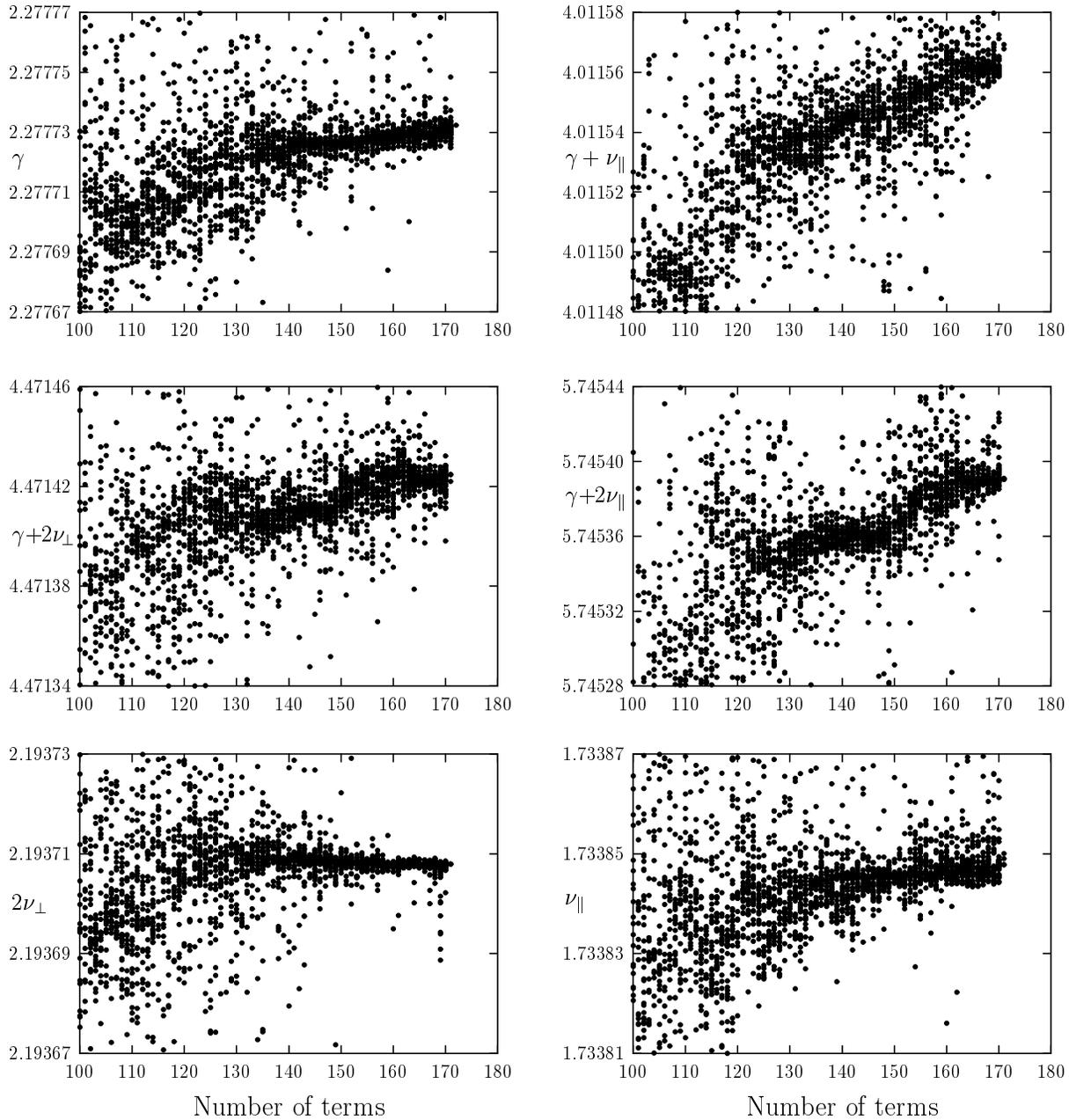}
\end{center}

\caption{\label{fig:ntexp} Estimates of the critical exponents
obtained from third-order differential approximants vs 
the number of terms used by the approximant. 
Shown are (from left to right and top to bottom) estimates from
the series $S(p)$, $\mu_{0,1}(p)$, $\mu_{2,0}(p)$, $\mu_{0,2}(p)$,
$\mu_{2,0}(p)\mu_{0,2}(p)/(\mu_{0,1}(p))^2$,
and  $\mu_{0,2}(p)/\mu_{0,1}(p)$.}

\end{figure}

In order to gauge the effect of any systematic drift and lack of
convergence to the true critical values it is helpful to plot
the estimates of  $p_c$ and associated exponents obtained from 
approximants to the various series as a function of the number of 
terms. Plotted in Fig.~\ref{fig:ntcrp} are estimates for $p_c$ 
obtained from third-order differential approximants with $L$ 
increasing from 0 to 50 in steps of 5. The orders $N_i$ were chosen 
so that the difference between the order of the polynomials $Q_i$ 
never exceeded 1. Each point in the figure represents an estimate from 
a single approximant. From this study, two conclusions
are immediately obvious. First of all most of the series show
some drift in the estimate for $p_c$. In particular one notes
that the estimates obtained from $\mu_{0,1}(p)$ and $\mu_{0,2}(p)$
(shown in the upper two panels on the right) do not appear to
have converged yet. The estimates from the remaining series largely
seem to have converged to a narrow band as the number of terms
exceed 150. Secondly it appears that generally the scatter among the 
various estimates becomes smaller as the number of terms increases.
All in all it appears that the estimates converge towards
$p_c \simeq 0.644700185$. The only notable exception is perhaps
the estimates from the series $\mu_{2,0}(p)\mu_{0,2}(p)/(\mu_{0,1}(p))^2$
(shown in the lower left panel) which seems to favor a slightly larger
value $p_c \simeq 0.644700191$, though in this case there still seems
to be a downwards trend in the estimates. It seems reasonable to
estimate that $p_c = 0.644700185(5)$. This is in good agreement
with the estimate $p_c = 0.64470015(5)$ obtained  previously
\cite{Jensen96a}. The new refined estimate is an order of magnitude 
more accurate and the central estimate lies within the error bounds
of the earlier estimate.
In Fig.~\ref{fig:ntexp} I have plotted estimates for the corresponding
critical exponents. Similar trends are apparent in this case. 
From these plots I venture the following estimates

\begin{eqnarray}\label{eq:expest}
\gamma  &= & 2.277730(5),\nonumber\\ 
\nu_{\parallel} & = & 1.733847(6), \nonumber\\
2\nu_{\perp}  &= & 2.193708(4), \nonumber\\
\gamma+\nu_{\parallel} & = & 4.01156(1), \\
\gamma+2\nu_{\parallel} & = & 5.74539(1), \nonumber\\
\gamma+2\nu_{\perp}  &= & 4.471425(15). \nonumber
\end{eqnarray}

\noindent
These estimates are again in full agreement with those
obtained previously, an order of magnitude or so more accurate, 
and while the central estimates again are trending higher they
still lie well within the earlier error bounds. Looking at
differential approximants of different order (second and fourth)
confirms the validity of these estimates.

I also analysed the series in order to estimate the leading confluent
exponents $\Delta_1$. As was the case for the percolation probability
series both the Baker-Hunter transformation and the method of
Adler, Moshe and Privman (see \cite{JGSQPP} and references therein
for details regarding these methods) yielded estimates consistent
with $\Delta_1 =1$. So there are no signs of non-analytic corrections
to scaling.

\subsection{The site series}

An analysis similar to that of the previous section was performed for
the site series. Generally I found that the estimates for $p_c$ were not 
so well converged. They seemed to change quite a bit as the order of
the approximant or inhomogeneous polynomial increased, and furthermore
the estimates also showed some inconsistencies from series to series.
The series for $\mu_{0,2}(p)$ and $\mu_{0,1}(p)$ yielded estimates
for $p_c \simeq 0.7054850$ while the series 
$\mu_{2,0}(p)\mu_{0,2}(p)/(\mu_{0,1}(p))^2$ favored an estimate
$p_c \simeq 0.7054853$ with the remaining series yielding estimates
in between. From these one could surmise that  $p_c=0.70548515(20)$.
The associated estimate for the critical exponents are listed below 

\begin{eqnarray}\label{eq:siteexpest}
\gamma  &= & 2.27765(6),\nonumber\\ 
\nu_{\parallel} & = & 1.73381(4), \nonumber\\
2\nu_{\perp}  &= & 2.19377(5), \nonumber\\
\gamma+\nu_{\parallel} & = & 4.01135(15), \\
\gamma+2\nu_{\parallel} & = & 5.7450(2), \nonumber\\
\gamma+2\nu_{\perp}  &= & 4.47130(8). \nonumber
\end{eqnarray}

\noindent
As can be seen these estimates are often only marginally consistent
with those from the bond series and at least an order of magnitude
less accurate. 

\begin{figure}
\begin{center}
\includegraphics{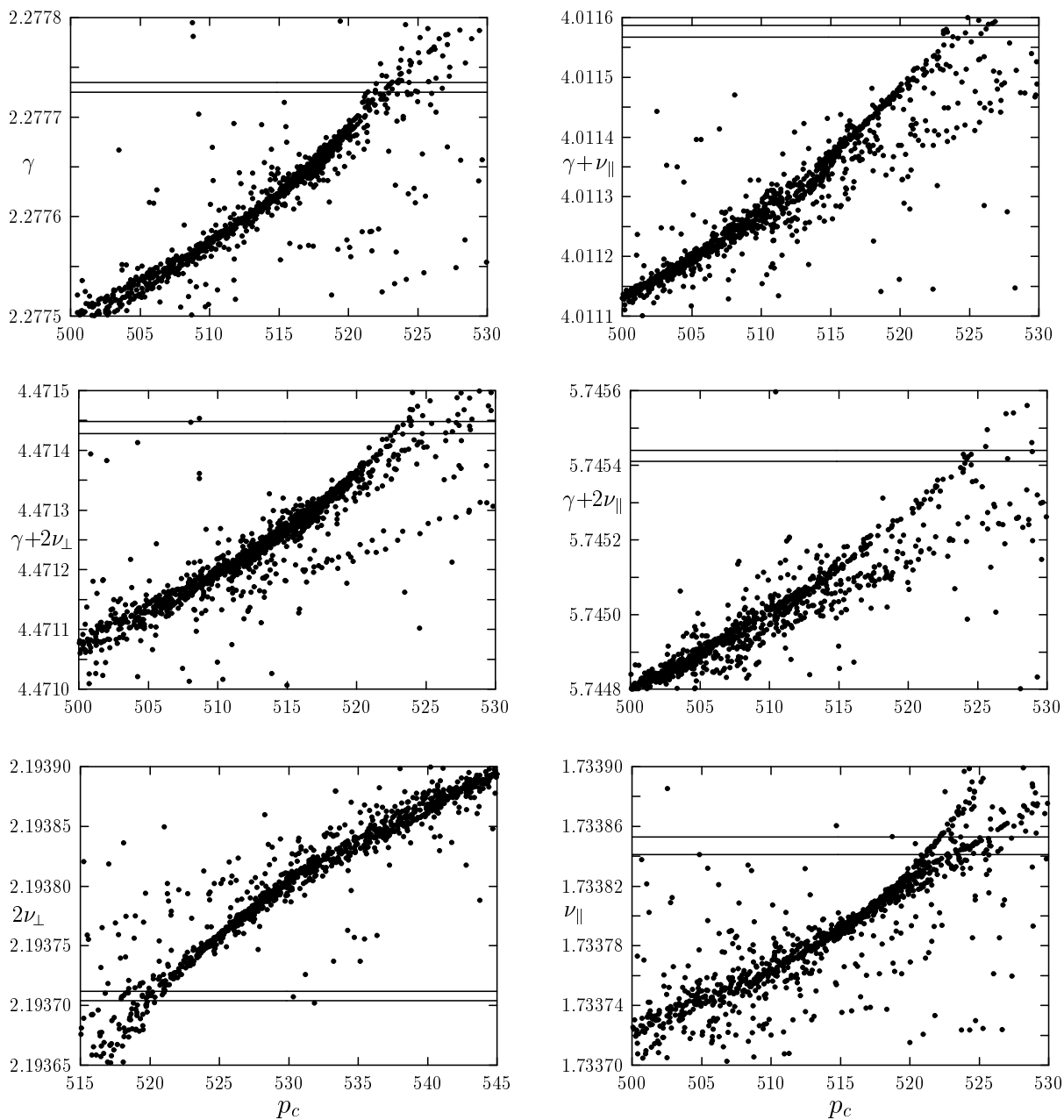}
\end{center}

\caption{\label{fig:sqscrpexp} Estimates of the critical exponents
obtained from third-order differential approximants vs
estimates of the critical point for square site problem. 
Numbers along the $x$-axis are all preceded by 0.70548.
Shown are (from left to right and top to bottom) estimates from
the series $S(p)$, $\mu_{0,1}(p)$, $\mu_{2,0}(p)$, $\mu_{0,2}(p)$,
$\mu_{2,0}(p)\mu_{0,2}(p)/(\mu_{0,1}(p))^2$,
and  $\mu_{0,2}(p)/\mu_{0,1}(p)$.}

\end{figure}

Due to the high degree of internal consistency of
the estimates from the bond series one would tend to believe
quite firmly in their accuracy and correctness and one can then
use them to try and obtain a more accurate estimate for $p_c$ in
the site problem. In Fig.~\ref{fig:sqscrpexp} I have plotted
the estimates for the critical exponents vs the estimates for $p_c$.
The solid lines indicate the error-bounds on the exponent estimates 
obtained by using the bond-series estimates for the exponents
$\gamma$, $\nu_{\parallel}$ and $\nu_{\perp}$. By extrapolating
the exponent estimates until they lie between the solid lines
it is obvious that this procedure yields a $p_c$ estimate consistent
with $p_c=0.70548522(4)$, which I take as the final estimate. 

\subsection{Non-physical singularities}

Non-physical singularities are of interest both because knowlegde
about their position and associated exponents may help in the search
for exact solutions and because one may gain a better understanding
of the problem by studying the behaviour of various physical 
quantities as analytic functions of complex variables. While 
comparatively little work has been done along these lines for
directed percolation this is a quite active field of study 
for classical spin systems such as the Potts and Ising models 
(see \cite{Shrock} for recent results and references to earlier work) .
  
The series have a radius of convergence smaller than $p_c$ due to 
singularities in the complex $p$-plane closer to the origin than the 
physical critical point. Since all the coefficients in the expansion 
are real, complex singularities always come in conjugate pairs. 
In order to locate the non-physical singularities in a systematic 
fashion I used the following procedure: Calculate all differential 
approximants with $K$ and $L$ fixed using at least 150 or 140 
terms in the bond or site cases, respectively. Each approximant yields 
$N_K$ possible singularities (and associated exponents) from the zeros 
of $Q_K$ (many of these are of course not actual singularities of the 
series). Next sort these `singularities' into equivalence classes by 
the criterion that they lie at most a distance $2^{-k}$ apart. An 
equivalence class is accepted as a singularity if it contains more than 
$N_a$ approximants ($N_a$ can be adjusted but I typically used a value
around 2/3 of the total number of approximants), and an estimate for 
the singularity and exponent is obtained by averaging over the 
approximants (the spread among the approximants is also calculated). 
This calculation was then repeated for $k-1$, $k-2$, $\ldots$ until a 
minimal value of 6. To avoid outputting well converged singularities 
at every level, once an equivalence class has been accepted, the 
approximants which are members of it are removed, and the subsequent
analysis is carried out only on the remaining data. This procedure
is applied to each series in turn, producing tables of possible
singularities. 

\begin{figure}
\begin{center}
\includegraphics{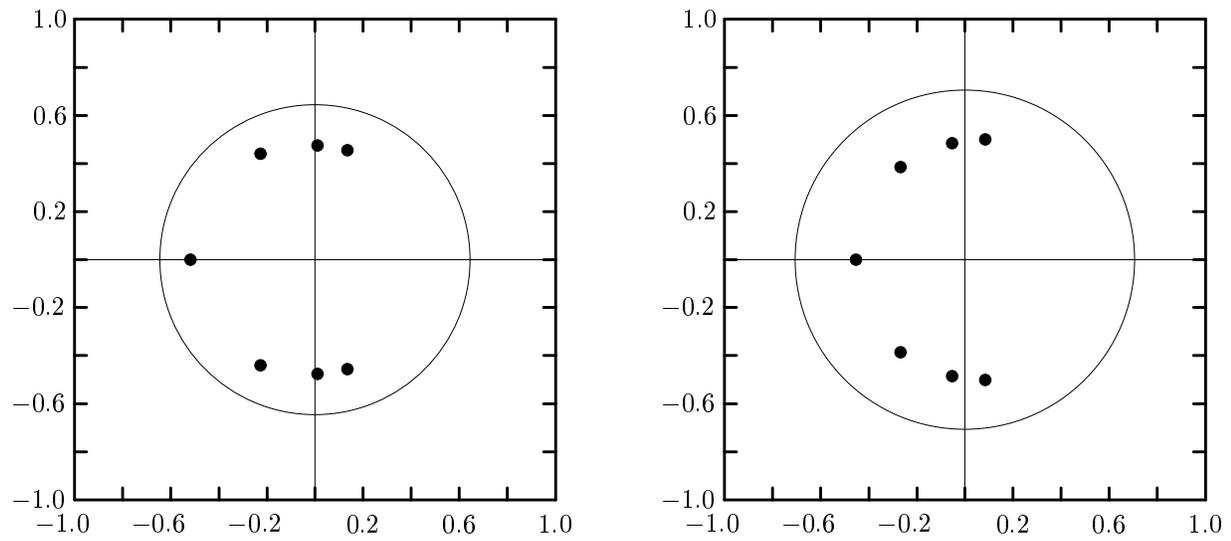}
\end{center}

\caption{\label{fig:npsing} Location of non-physical singularities
for bond (left panel) and site (right panel) directed percolation.}

\end{figure}

The analysis indicates that the series have quite a large number of 
non-physical singularities. A quick view of the distribution of
singularities can be gained from Fig.~\ref{fig:npsing} which show
the location of the non-physical singularities. Estimates for the
non-physical singularities are listed in Table~\ref{table:npsing}.
For both the bond and site cases there are two sets of singularities.
For those marked by 1, quite accurate estimates can be obtained for
both the location of the singularity and the associated exponents.
The exponent estimates for the site problem at the singularity on the
negative axis are consistent with the exact values 1/2, $-1/2$, $-3/2$, 
and $-1/2$ for the series $S(p)$, $\mu_{0,1}$, $\mu_{0,2}$, and 
$\mu_{2,0}$, respectively. At the conjugate pair of complex singularities 
the corresponding exponents are consistent with the exact values 3, 2, 1 
and 2, respectively. In either case the error is no more than 
0.1\%. In the bond case the exponents at the conjugate pair of complex
singularities seem to be identical to those for the site problem
though the error is at least an order of magnitude larger. At the 
singularity on negative axis the exponent estimates are $-0.075(5)$,  
$-1.05(2)$,  $-2.025(15)$, and $-1.00(2)$. So the most likely scenario 
for exact values would seem to be a logarithmic divergence in S(p), 
and divergences with exponents  $-1$, $-2$, and $-1$ in  $\mu_{0,1}$, 
$\mu_{0,2}$, and $\mu_{2,0}$, respectively.
For the singularities marked 2, the estimates for the location of the
singularity is poor and no meaningful estimates can be obtained for
the exponents. In all cases the singularities are found in all
series and remain fairly stable as $K$ and $L$ is varied.

\section{Conclusion \label{sec:conc}}

In this paper I have reported on a new algorithm for the derivation of
low-density series for moments of the pair-connectedness in directed
percolation problems. Numerical evidence shows that the 
computational complexity grows exponentially, but with a growth factor 
$\lambda < \sqrt[8]{2}$, which is much smaller than the growth 
factor $\lambda = \sqrt[4]{2}$ of the previous best algorithm.
For bond (site) percolation on the directed square lattice the series 
have been extended to order 171 (158) as compared to order 112 (106) 
obtained in previous work \cite{Jensen96a}.

Analysis of the bond series led to a very accurate estimate for the 
critical point,  $p_c = 0.644700185(5)$, and the values of the critical 
exponents for the average cluster size, parallel and perpendicular 
connectedness lengths are estimated to be $\gamma  =  2.277730(5)$, 
$\nu_{\parallel}  = 1.733847(6)$ and $\nu_{\perp}  =  1.096854(4)$. 
An accurate estimate for the percolation probability exponent is 
obtained from the scaling relation
$\beta = (\nu_{\parallel}+\nu_{\perp}-\gamma)/2 = 0.276486(8)$. 

For reference purposes I  list  estimates for some other
critical exponents obtained using various scaling relations:
\begin{eqnarray*}
\Delta = & \beta+\gamma = & 2.554216(13) \\
\tau = & \nu_{\parallel}-\beta = & 1.457362(14) \\
z = & \nu_{\parallel}/\nu_{\perp}  = & 1.580745(10) \\
\gamma ' = & \gamma-\nu_{\parallel} = & 0.543883(11) \\
\delta = & \beta/\nu_{\parallel} = & 0.159464(6) \\
\eta = & \gamma/\nu_{\parallel}-1 = & 0.313686(8). 
\end{eqnarray*}
Here $\Delta$ is the exponent characterizing the scale of the cluster
size distribution, $\tau$ is the cluster length exponent, $z$ is the
dynamical critical exponent, $\gamma '$ the exponent characterizing
the steady-state fluctuations of the order parameter, while $\delta$ 
and $\eta$ characterize the behaviour at $p_c$ as $t\rightarrow \infty$
of the survival 
probability and average number of particles, respectively.

Assuming that the exponent estimates from the square bond case
are correct, an improved $p_c$-estimate was obtained for the
square site problem, $p_c=0.70548522(4)$.

Finally I note, that an analysis of the series, in order to
determine the value of the confluent exponent, yielded estimates
consistent with $\Delta_1 \simeq 1$. Thus there is no evidence of
non-analytic corrections to scaling. 

\section*{E-mail or WWW retrieval of series}

The series for the directed percolation problems on the various lattices 
can be obtained via e-mail by sending a request to 
I.Jensen@ms.unimelb.edu.au or via the world wide web on the URL
http://www.ms.unimelb.edu.au/\~{ }iwan/ by following the instructions.

\section*{Acknowledgments}
 
I would like to thank A. Guttmann for many useful comments on the
manuscript and the Department of Computer Science for very generous
allocations of computing resources.
The work was supported by a grant from the Australian Research Council.

\clearpage

\begin{table}
\caption{\label{table:npsing}
Estimates of the location of various non-physical singularities
in the series for directed bond and site percolation.}
\begin{center}
\begin{tabular}{cll}  \hline  \hline
 & Bond & Site \\ \hline
1 & $-0.51670(3)$ & $-0.451952165(8)$ \\
1 & $-0.22605(10)\pm 0.4400(1)i$ & $-0.2661783(4)\pm 0.3847813(4)i$ \\
2 &  $0.135(10)\pm 0.455(10)i$ & $-0.0525(5)\pm 0.4840(5)i$ \\
2 &  $0.0105(10)\pm 0.475(1)i$ & $0.085(15)\pm 0.500(15)i$\\
 \hline  \hline
\end{tabular}
\end{center}
\end{table}

\clearpage

\end{document}